# Performance of Graduate Students at Identifying Introductory Physics Students' Difficulties Related to Kinematics Graphs

Alexandru Maries and Chandralekha Singh

*Department of Physics and Astronomy, University of Pittsburgh, Pittsburgh, PA 15260*

**Abstract.** The Test of Understanding Graphs in Kinematics (TUG-K) is a multiple choice test developed by Beichner to assess students' understanding of kinematics graphs. Many of the items on the TUG-K have strong distractor choices which correspond to introductory students' common difficulties with kinematics graphs. Instruction is unlikely to be effective if instructors do not know these common difficulties and take them into account in their instructional design. We evaluate the performance of first year physics graduate students at identifying introductory students' common difficulties related to kinematics graphs. In particular, for each item on the TUG-K, the graduate students were asked to identify which incorrect answer choice they thought would be most commonly selected by introductory physics students if they did not know the correct answer after instruction in relevant concepts. We used the introductory student data from Beichner's original paper to assess graduate students' knowledge of introductory students' difficulties. Furthermore, we selected the four questions on the TUG-K on which the graduate student performance was the poorest for a more detailed analysis which included think-aloud interviews. We present results which can partly account for the poor graduate student performance on these questions and also inform instruction in professional development courses.

## INTRODUCTION

The Test of Understanding Graphs in Kinematics (TUG-K) [1] is a multiple choice test developed by Beichner in 1994 which takes into account the common difficulties of introductory students in constructing and interpreting kinematics graphs identified by previous research [2,3]. It reveals that many introductory students have difficulties with kinematics graphs even after instruction in relevant concepts. The research presented here used the TUG-K to explore one aspect of the pedagogical content knowledge (PCK) of physics graduate students (GSs), namely, their knowledge of common introductory student difficulties related to kinematics graphs, which is an important aspect of PCK as discussed, e.g., by Shulman [4]. Knowledge of these difficulties and of the types of reasoning used by introductory students can be helpful in designing pedagogical strategies to improve their learning [4,5].

## RESEARCH QUESTIONS

Here, we focus on three research questions:
**RQ1.** How do American physics GSs, who have been exposed to undergraduate teaching in the United States, compare to foreign GSs in their performance at identifying introductory student difficulties?

According to recent statistics, roughly half of all physics GSs across the United States are foreign. Since a majority of GSs serve as teaching assistants (TAs) for at least one or two semesters, the impact they have on undergraduate education is considerable and comparable to that of American GSs. It is therefore useful to compare the knowledge that American GSs and foreign GSs have about introductory student difficulties.
**RQ2.** To what extent do GSs identify "major" student difficulties compared to "moderate" ones?

Introductory students' difficulties related to kinematics graphs are of varying degrees, and while one might assume that the more common difficulties are easier to identify, this may not be the case. We therefore investigated whether there is a difference in the extent to which GSs identify "major" vs. "moderate" difficulties.
**RQ3**. What major common introductory student difficulties are identified by very few GSs and why?

Since GSs can play a significant role in improving introductory student understanding, it is useful to determine which major introductory student difficulties most GSs are unable to identify and what prevents them from being able to identify them. This knowledge can be useful in, e.g., appropriately designing interventions for professional development of TAs.

## METHODOLOGY

The participants in this study [6] were twenty-five first year physics GSs, most of whom were serving as TAs. The GSs were enrolled in a semester-long TA training class. The nationalities of the TAs varied, nine of them being American, nine Chinese and seven from other countries. The materials used in this investigation include the TUG-K and the introductory student post-test data collected by Beichner for his study, which were from more than 500 college and high-school students.

Towards the end of the semester (so that each TA had one semester of classroom teaching experience), as part of a two hour long TA training class, GSs completed the following TUG-K related tasks:
1. Individually, for each question, the GSs identified the correct answer.
2. Individually, for each question, the GSs identified which one of the four *incorrect* answer choices in their view, would be *most commonly selected by introductory physics students* after instruction in relevant concepts. This will be referred to as the *TUG-K related PCK task* here.

The TUG-K related PCK task was framed as a task in which GSs had to identify introductory students' difficulties after students had traditional instruction in relevant concepts (as opposed to before instruction) because many GSs noted that they have no way of knowing what students would select before instruction. Physics education research has often revealed that student difficulties remain similar after instruction, except that fewer students have those difficulties. Therefore, framing the TUG-K related PCK task to be for a post-test (after introductory students had traditional instruction in relevant concepts) rather than a pre-test provides similar information about GSs knowledge of introductory students' difficulties, with the added benefit that it makes the task easier to complete for GSs.

In order to quantify the TUG-K related PCK performance of GSs, scores were assigned to each graduate student for each question [6]. A graduate student who selected a particular incorrect answer choice as the most common incorrect choice for a particular question received a PCK score which was equal to the fraction of introductory students who selected that particular incorrect answer choice. For example, for Question 1 on the TUG-K, the fractions of introductory students who selected choices A, B, C, D and E (Table III in Ref. [1]) are 0.41, 0.16, 0.04, 0.22 and 0.17, respectively. Answer choice B is correct, thus, the PCK score assigned to a graduate student for each answer choice if he/she selected it as the most common incorrect answer would be 0.41, 0, 0.04, 0.22 and 0.17 (A, B, C, D and E). The total PCK score a graduate student would obtain on the task for the entire TUG-K can be obtained by summing over his/her scores for each of the 21 TUG-K questions. We note that our approach used to determine the TUG-K related PCK score weighs the responses of GSs by the fraction of introductory students who selected a particular incorrect response in order to have the scores received by GSs on a particular question be representative of the prevalence of the introductory student difficulties they identified.

In addition to the quantitative data, qualitative data were collected via semi-structured audio-recorded individual think-aloud interviews. Six GSs who were not from the class in which the group study was carried out but were TAs in at least one previous semester volunteered to be interviewed. During the interviews, they completed the TUG-K related PCK task for the four questions in which the quantitative study indicated that GSs on average had the lowest performance in identifying student difficulties. They were not disturbed during the think-aloud process. Only at the end, we asked them for clarification of the points they had not made clear on their own and probed them further to understand their reasoning.

## RESULTS

GSs in both the TA training class and in the interviews explicitly noted that the task of thinking from an introductory student's point of view was challenging. During the discussions, they also articulated that they did not expect introductory students to have certain difficulties, and that sometimes they were surprised by the high percentage of students having those difficulties.

**RQ1.** The best possible PCK score on the TUG-K related PCK task would correspond to selecting the most common incorrect answer choice for each question. This would amount to a score equal to the sum of the largest fractions of incorrect answer choices for each question, which turns out to be 6.70. Table 1 shows the averages and standard deviations of the TUG-K related PCK scores of three different groups of GSs. Chinese GSs were placed in a separate group because they comprise more than half of the foreign graduate student population, comparable to the number of American students. A non-parametric test (Kruskal-Wallis) reveals that the three groups exhibit comparable performance on the TUG-K related PCK task. Also, the standard deviations are small. Thus, it appears that American GSs and foreign GSs display comparable performance in identifying student difficulties on the TUG-K.

**TABLE 1.** Numbers of American/Chinese/Other foreign GSs (N), their averages (and percentage of those averages out of the maximum PCK score) and standard deviations (Std. dev.) for the PCK scores obtained for determining introductory student difficulties.

|  | N | Average | Std. dev. |
|---|---|---|---|
| American | 9 | 4.00 (60%) | 0.54 |
| Chinese | 9 | 4.24 (63%) | 0.55 |
| Other foreign | 7 | 4.46 (66%) | 0.59 |

**RQ2.** Two researchers mutually agreed on a heuristic that a "major" student difficulty was related to an incorrect answer choice selected by more than 1/3 of the introductory students and a "moderate" difficulty was related to an incorrect answer choice selected by between 1/5 and 1/3 of introductory students. There are 17 questions which fit either of these two criteria, eight of which are related to major introductory student difficulties and nine of which are related to moderate student difficulties (only two questions, 1 and 16, have both a major and moderate difficulty, and they were categorized as questions with major difficulties). Table 2 shows that GSs performed worse at identifying introductory physics students' major difficulties compared to moderate ones.

**TABLE 2.** GSs' average TUG-K related PCK score (as a percentage of the maximum possible) on questions with major and moderate difficulties.

| Major difficulties | 48% |
|---|---|
| Moderate difficulties | 61% |

Furthermore, Table 3 shows all the questions in which the GSs' TUG-K related PCK score was lowest (less than 30% of the maximum possible score). These four questions are all related to major difficulties. Thus, in half of the questions in which introductory students had major difficulties, the GSs had the lowest PCK score at identifying these difficulties.

**TABLE 3.** Questions in which the GSs average TUG-K related PCK score is less than 30%.

| Question # | 6 | 8 | 9 | 17 |
|---|---|---|---|---|
| GSs performance | 28% | 22% | 29% | 28% |

**RQ3.** Here, we only focus on the questions with the lowest GSs' PCK score at identifying introductory student difficulties shown in Table 3. GSs' reasoning for their choices for these questions was probed during the semi-structured think-aloud interviews.

Questions 6 and 17 both require students to compute slopes, and Table 4 shows that roughly half of introductory students do not examine initial conditions carefully (i.e., they compute the slope as $y/x$ instead of $\Delta y/\Delta x$ as discussed in Ref. [1]). These two questions are the only ones which involve this difficulty on the TUG-K. A less common introductory student difficulty with computing slopes is confusing the slope with the ordinate value at the corresponding abscissa (i.e., reading off the $y$ value). Table 4 shows that few GSs (20% and 16%) identified the major difficulty of roughly half of the introductory students, but more GSs identified the less common introductory student difficulty. Table 4 also provides support to the data in Table 2 to answer RQ2 in a particular context.

During the interviews, for Questions 6 and 17, all GSs initially approached the TUG-K related PCK task by working "forward": they started by thinking about the incorrect types of reasoning introductory students could employ. Then, they tried to identify if these incorrect types of reasoning led to one of the incorrect answer choices given in TUG-K. Typically, these attempts did not lead the GSs to an incorrect answer choice for either of these questions (which indicates that GSs were unaware of the corresponding common introductory student difficulties). For example, Tony said: "I'm trying to think about what they, [short pause], what could go wrong here", and a while later after trying a few incorrect approaches which didn't lead to any distractors in TUG-K he said "I don't know, I guess I would probably be surprised by what could go wrong here […] I can't think of [what students could do]".

**TABLE 4.** The two common introductory student difficulties with computing slopes, the percentage of introductory students who had those difficulties, and the percentage of GSs who identified these difficulties as the most common.

| Q # | Introductory students | | GSs | |
|---|---|---|---|---|
| | Slope=$y/x$ | Slope=$y$ value | Slope=$y/x$ | Slope=$y$ value |
| 6 | 46% | 16% | 20% | 36% |
| 17 | 46% | 19% | 16% | 44% |

Similar to Tony, after the first few unsuccessful attempts, all interviewed GSs would typically start working "backwards", i.e., they looked at the distractors given and tried to deduce why introductory students may select them. Since one of the distractors in each question was the $y$ value at the $x$ value in the region where students were supposed to compute the slope, most interviewed GSs readily selected it as the most common incorrect answer choice. For example, after noticing the distractor Michelle said "I think [this distractor] might be … [the most common] incorrect answer because they'll basically say 'ok, I need to find it at [$x$ value], ok, it's just [the corresponding $y$ value]. Maybe they'll just read off the graph like that." After identifying this difficulty as most common, Michelle or other GSs did not carefully consider the other distractors and try to articulate the incorrect reasoning that would lead introductory students to select them. Therefore, the interviewer explicitly asked the GSs to deduce the student reasoning that would lead the students to select the incorrect answers which were related to the "slope as $y/x$" difficulty. Most interviewed GSs were unable to do so which provides further evidence that the GSs had not thought about this common difficulty of introductory students. In addition, in the final phase of the interview when common introductory student difficulties were discussed, several GSs explicitly noted that they were surprised that introductory students would compute the slope as $y/x$. For example, Tyler said: "I was pretty surprised by the fact that they're just doing y over x [...] I would think that the students that are realizing that y over x is the slope would realize that it is delta y over delta x. The fact that they're getting that far, but not quite [getting it] is surprising to me."

Questions 8 and 9 both test students' ability to translate between two representations of motion, verbal and graphical. In both of these questions, the major introductory student difficulty was to match the form of the graph with the verbal description superficially, without careful examination of the $y$ axis. For example, question 8 provides a three part graph of displacement vs. time (positive and constant, then steadily decreasing, then zero) and asks students for the correct verbal interpretation. The most common incorrect introductory student choice treats the *y* axis as velocity rather than displacement (i.e., this interpretation would be correct if the graph was of velocity vs. time because the answer choice states that the object first moves at a constant velocity, then slows down, and then stops). Identical incorrect reasoning for the displacement vs. time graph leads to the most common introductory student incorrect answer choice in Question 9. Table 5 shows that on both of these questions, few GSs (8% and 28%) identified this major student difficulty.

Instead, for question 8, all interviewed GSs claimed that introductory students would interpret the steadily decreasing part of the position vs. time graph as a "hill", and therefore all of them selected as the most common incorrect answer of introductory students one of the three answer choices which included a statement that the object rolls or moves down a hill. For example, Tyler said: "I could see choice E getting picked a lot because basically if you just look at it, it's a flat area and then rolling down a hill, and then it keeps moving, so if you just imagine rolling a ball along this curve, that's what would happen." However, Beichner's data [1] indicates that, on this question, this difficulty is less common than the difficulty of matching the graph with the verbal description superficially without regard for the axes (for example, only 5% of introductory students selected "E", the incorrect answer choice favored by Tyler).

**TABLE 5.** Percentage of introductory students who had the major difficulty discussed on Questions 8 and 9, and percentage of GSs (GSs%) who identified this difficulty as the most common one.

| Q # | Intro. student difficulty % | GSs % |
|---|---|---|
| 8 | 37% | 8% |
| 9 | 57% | 28% |

## DISCUSSION AND SUMMARY

Understanding how introductory students reason about physics is important for instructors and GSs in order to take advantage of introductory students' initial knowledge, and develop and adapt effective pedagogical approaches. This investigation used the TUG-K to evaluate the knowledge that GSs have about common student difficulties related to kinematics graphs.

We find that American GSs who had been exposed to undergraduate teaching in the US exhibit similar performance to foreign GSs at identifying common introductory student difficulties on the TUG-K. The discussion in the TA training class after the GSs completed the TUG-K related PCK task and interviews also suggest that the foreign GSs were similar to American GSs in their thought processes in this regard. We note that this study was not designed to probe why these groups exhibited comparable PCK performance when identifying common introductory student difficulties with kinematic graphs despite their different backgrounds.

We also find that on average, GSs identified major introductory student difficulties less often than moderate ones. Furthermore, the analysis of the PCK score of the GSs (as a percentage of the maximum possible score) shows that the four questions on which the average PCK score of GSs was the lowest were all related to major introductory student difficulties.

A detailed analysis of the PCK scores of GSs on those four questions revealed that in Questions 6 and 17 GSs identified the less common introductory student difficulty related to computing slopes (reading off the *y* value) more often than the more common difficulty (computing the slope as *y/x*). Individual interviews suggest that GSs struggle to identify the most common difficulty even after being explicitly asked to determine how introductory students could select the answer choices connected to the "slope as *y/x*" difficulty. Furthermore, most GSs struggled to identify the common introductory student difficulty of matching a graph with a verbal description superficially, without regard for the axes (Questions 8 and 9). Interviews suggest that on Question 8, this was the case for all interviewed GSs who claimed that introductory students would interpret the sloping part of the graph as a hill (which was not the most common difficulty).

Finally, both in the interviews and in the TA training class, many GSs remarked that the task of thinking from the point of view of an introductory student was challenging. However, the discussion which followed the TUG-K related PCK task suggested that they found it useful because it made them more aware of the patterns of reasoning of introductory physics students. In addition, interviewed GSs noted that identifying the most common incorrect answer choices on other assessments designed by PER researchers would be very useful for professional development programs for TAs and new faculty if a discussion of common student difficulties ensues after performing the task. For example, when asked if one should use such tasks in a physics teaching related professional development course or workshop, Tyler remarked: "I think so, yea. I think it would definitely be useful cause […] if you don't know what their misconceptions are, how are you going to correct them?"